\documentclass{ws-procs975x65}

\def\beq{\begin{equation}}
\def\eeq{\end{equation}}

\begin{document}
\title{Motion of  test  particles in a magnetized conformastatic   background}
\author{Antonio C. Guti\'errez-Pi\~{n}eres}

\address{Facultad de Ciencias B\'asicas, Universidad Tecnol\'ogica de Bol\'ivar, \\ Cartagena 13001, Colombia\\
Instituto de Ciencias Nucleares, Universidad Nacional Aut\'onoma de M\'exico, \\
AP 70543,  M\'exico, DF 04510, M\'exico\\
E-mail: gutierrezpac@gmail.com}

\author{Abra\~{a}o J. S.  Capistrano}

\address{Federal University of  Latin-American Integration, \\
Casimiro Montenegro Filho Astronomical Pole, Technological Park of Itaipu\\
Technological Park of Itaipu, PO box 2123, Foz do Igua\c{c}u-PR 85867-670, Brazil\\
E-mail: abraao.capistrano@unila.edu.br}

\begin{abstract}
A class of  exact   conformastatic  solutions of  the Einstein-Maxwell field equations is  presented in which the gravitational and electromagnetic
potentials are completely determined by a harmonic function only.  The  motion of  test  particles is  investigated  in the
background of a  space-time  characterized  by  this  class of solutions.  We focus on the study of circular stable and unstable orbits  obtained by
taking  account particular harmonic  functions defining the  gravitational potential.  We  show that is possible to  have  repulsive  force  generated
by the  charge  distribution of  the  source.  As the  space-time  here  considered  is   singularity  free we  conclude that  this  phenomena  is
not  exclusive to the  case  of  naked  singularities. Additionally,  we  obtain an  expression  for   the  perihelion advance  of the test  particles
in  a  general   magnetized  conformastatic  space-time.
\end{abstract}

\bodymatter


\section{A conformastic  solution of  the  Einstein-Maxwell equations}
\label{sec:Introduction}
In Einstein-Maxwell gravity  theory  the  background of  a  symmetric  body  can  be  described  by the 
 conformastatic metric in cylindrical coordinates  \cite{PhysRevD.87.044010}
\begin{equation}
      dS^2 = - c^2 e^{2\phi} dt^2 + e^{-2\phi} (dr^2 + dz^2 + r^2d\varphi^2), \label{eq:metric}
\end{equation}
where  $c$ is the  speed of  the  light in the  vacuum   and the  metric  potential $\phi$ depends only on the $r$  and  $z$ variables. The  governing
equations  field  are  the  Einstein-Maxwell equations which,  for  the  metric (\ref{eq:metric}),  can be  equivalently written  as 
\footnote{Along this  work  we  use the  CGS units such that  $k_{_0} \equiv 8\pi\, G c^{-4} =2,07\times10^{-48}$
s$^2$cm$^{-1}$g$^{-1}$,  $G= 6.674\times10^{-8}$cm$^{3}$g$^{-1}$s$^{-2}$
and  $c=2.998\times 10^{10}$\,cm\,s$^{-1}$.},
\begin{eqnarray}
  \nabla^2 \phi     = \nabla \phi \cdot  \nabla \phi,\label{eq:EMexpli1}\quad
  \phi_{,r}^{\,2}   = \frac{G }{c^4 r^2} e^{2\phi} A_{\varphi,z}^2,\quad
  \phi_{,z}^{\,2}   = \frac{G }{c^4 r^2} e^{2\phi} A_{\varphi,r}^2,\\
  \phi_{,r}\phi_{,z}=-\frac{G }{c^4 r^2} e^{2\phi} A_{\varphi,r}A_{\varphi,z},\quad
  \nabla \cdot (r^{-2} e^{2\phi} \nabla{A_{\varphi}})=0,
\end{eqnarray}
where the  operator $\nabla$  denotes  the usual gradient  in  cylindrical  coordinates.  By  supposing a  solution of $\nabla^2 \phi     = \nabla \phi \cdot  \nabla \phi$
in the  functional  form $\phi=\phi[U(r,z)]$,  where $U(r,z)$ is  an arbitrary harmonic  function ( See Refs.~\refcite{PhysRevD.87.044010} and \refcite{GRG2015ACG-P2015} 
for more details) it is  not difficult  to prove that
\begin{equation}
 \phi=-\ln{(1 - U)}, \quad
  A_{\varphi}(r,z) = \frac{c^2}{G^{1/2}}{\int_0^r{\tilde{r} U({\tilde{r},z}}}) d \tilde{r},
 \label{eq:solution}
\end{equation}
  is a solution  of  the previous  equations  system. The electromagnetic field is of a magnetic type. In fact,  can be demonstrated  that  the 
 electromagnetic invariant ${\cal  F}\equiv F^{\alpha\beta} F_{\alpha\beta} \geq 0$.The  nonzero  components  of  the  electromagnetic  fields  
 read  $B_i= {c^2}{G^{-1/2}}rU_{,i} $,   $(i=r,z)$ where,  as we  know, $U$ 
is  an arbitrary  harmonic  function.

The  motion  of  a test  particle of mass $m$ and  charge $q$ moving  in a   conformastatic space-time  given  by  the  line element (\ref{eq:metric})
is  described  by the Lagrangian
  ${\cal L} =\frac{1}{2}m g_{\alpha\beta}  \dot{x}^{\alpha}\dot{x}^{\beta} + \frac{q}{c}A_{\alpha}\dot{x}^{\alpha},$
where  $A_{\alpha}$ is  the four-potential  and the  dot represents differentiation with  respect to the  proper time.  Since  the   Lagrangian does
not  depend  explicitly  of the  variables $t$ and $\varphi$,  the  following  two conserved  quantities  exist
 \begin{equation}
   p_t= -mc e^{2\phi}\dot{t} \equiv -\frac{E}{c},\quad \mbox{and}\quad
   p_{\varphi}= m r^2 e^{-2\phi}\dot{\varphi} + \frac{q}c{}A_{\varphi} \equiv L,
   \label{eq:angularmomentum}
\end{equation}
where $E$ and $L$ are, respectively,  the  energy and the  angular  momentum  of  the particle  as  measured by  an observer at  rest at infinity. The
 momentum $p_{\alpha}$ of  the  particle  can be  normalized  so that $ g_{\alpha\beta} \dot x^{\alpha} \dot x^{\beta} = - \Sigma, $ accordingly, for
the  metric
(\ref{eq:metric}) 
\begin{equation}
 -{c^2e^{2\phi} \dot{t}^2} + e^{-2\phi}(\dot{r}^2  + \dot{z}^2 + r^2 \dot{\varphi}^2  ) =- \Sigma,
 \label{eq:circuprecess}
\end{equation}
where $\Sigma = 0, c^2, -c^2$ for null,  time-like,  and  space-like curves, respectively.

To investigate the  motion  of  the  test  particle  we  restrict ourselves  to the  study  of the  equatorial trajectories with $z=0$. Accordingly,
by the  substitution of the  conserved quantities (\ref{eq:angularmomentum}) into Eq.(\ref{eq:circuprecess}) we  find
\begin{equation}
  \dot{r}^2 + \Phi = \frac{E^2}{m^2c^2}, 
\quad \mbox{where} \quad
  \Phi(r) \equiv \frac{L^2}{m^2r^2}\left(1 - \frac{qA_{\varphi}}{Lc}  \right)^2 e^{4\phi} + \Sigma e^{2\phi}
  \label{eq:effectivepot}
\end{equation}
is  an effective  potential.   In this    work   we  present an  analysis  of  the circular  motion of  test  particles governed  by  the  above  equations. 
As  a  particular sample we consider  the  effective  potential  corresponding  to  a charged particle moving  in the gravitational  field   of  a  punctual  source
placed  in the  coordinates  origin of   a magnetized  conformastatic  space-time.   Additionally,  we find    an  expression  for   the  perihelion
advance  of the test  particles in  a  general   magnetized conformastatic  space-time.   This  work  is  inspired  in  the  Refs.~\refcite{PhysRevD.83.104052}  
and  \refcite{PhysRevD.83.024021}.


\section{Circular motion   in the  field  of  a  punctual source in Einstein-Maxwell  gravity}
The    motion  of  charged test particles  is  governed  by  the  behavior of  the  effective potential (\ref{eq:effectivepot}). The  radius of
circular  orbits and  the corresponding values of  the  energy $E$ and  angular momentum  $L$ are  given by  the  extrema of the  function $\Phi$.
Therefore, the  conditions  for  the  occurrence of  circular  orbits  are
\begin{equation}
  \frac{d\Phi}{dr} =0, \quad  \Phi= \frac{E^2}{m^2c^2}.
  \label{eq:circularconditios}
\end{equation}
Notice  that the  positive  value  of  the  energy  corresponds to  the positive of solution $E_{\pm} =\pm m c\Phi^{1/2}$. 
An  interesting particular  orbit  is  the  one  in which  the  particle is  located  at  rest   $(r_r)$ as  seen  by an observer at  infinity,
i.e. $L=0$. These orbits are  therefore characterized  by  the  conditions
 $L=0, \quad {d \Phi}/{dr} =0.$
For  the metric (\ref{eq:metric})  these  conditions give  us  the  following  equation  for the rest  radius
\begin{equation}
\frac{2 e^{2 \phi}}{ m^2 c^2 r^3 } \left[  q^2 A_{\varphi}  \left(2 r  A_{\varphi} \phi_{,r}
            + r  A_{\varphi, r} -  A_{\varphi} \right) e^{2 \phi}
            + \Sigma  m^2c^2  r^3 \phi_{,r}\right] =0. \label{eq:rrest}
\end{equation}
If  $ e^{2 \phi}=0$ for  a  orbit
with a  rest radius $r=r_r$ the  energy  of  the  particle  is  $E_r= 0$. In the  case $ e^{2 \phi}
\neq 0$  we  have
\begin{equation}
 E_{\quad r  \pm}^{(\pm)} = \pm m c e^{\phi} \left(\Sigma + \xi^{(\pm)}_{\quad r} \right)^{1/2}
 \label{eq:energyrest},
\end{equation}
where
\begin{equation}
\xi^{(\pm)}_{\quad r} =\frac{q^2 e^{2\phi}   \left[  r A_{\varphi, r} \pm \left(  r A_{\varphi, r}
+ 2 A_{\varphi}  \left(2 r \phi_{,r} - 1\right)   \right)  \right]^2}
{4 m^2 c^2 r^2   \left(2 r \phi_{,r} - 1\right)^2},
\end{equation}
To  obtain  $r_r$  we  must  to  solve  the  equation (\ref{eq:rrest}) for  $r_r$.
The  minimum radius  for  a  stable circular  orbits  occurs  at  the  inflection  points of  the  effective potential function,  thus  we  must
solve  the  equation
  ${d^2\Phi}/{dr^2} =0,$
using    the  expresion of  the angular moment $L$ of  the  particle  in  a  circular  orbit.   From  the equation (\ref{eq:circularconditios}) and
the last  equation we  find  that  the  radius of  last stable circular orbit and  the  angular moment of  this orbit. 

 The gravitational  and  magnetic  field in the  background  of a source  of  mass  $M$  placed  in the  origin  of  coordinates in  Einstein-Maxwell  gravity
 can be  obtained by  consider   the harmonic  potential
\begin{equation}
    U(r,z)=-\frac{G M }{c^2 R},  \quad R^2= r^2 + z^2
    \label{eq:Upotential}
\end{equation}
in the  equation (\ref{eq:solution}). Accordingly,  for  the  metric  and  electromagnetic  potential  we have
\begin{equation}
\phi(r, z)= - \ln{  \left(1 + \frac{G M}{c^2 R}  \right) }
\quad  \mbox{and} \quad
A_{\varphi}(r, z)= \sqrt{G} M \left( 1 - \frac{z}{R} \right)
\label{eq:explicsolmag}
\end{equation}
respectivly.  By  calculating  the  asymptotic  behavior  of  the metric  potential $g_{tt}$,  the Kretschmann scalar and  the electromagnetic 
invariant $\cal F$  we
conclude  that the  gravitational  field  is asymptotically like Schwarzschild  and  singularity free.  We  notice that when  we have the  solution Eq.(\ref{eq:Upotential}),
  we  recover  the  extreme Reissner-Nordstr\"{o}m solution  in  its  standard   form  by  writing  the metric in spherical  polar  coordinates  and
replacing   $R$ by  $R - GM/c^2$.

Consider  the case  of  a  charged  particle moving  in the   fields   given   by 
  (\ref{eq:explicsolmag}).
This  means  that  we  are considering  the  motion  described  by  the  following  effective  potential
\begin{equation}
 \Phi(r)= \frac{c^6 r^2 (Lc - q\sqrt{G}M)^2}{m^2 (c^2r + G M)^4}
           + \frac{\Sigma c^4 r^2}{(c^2r + G M)^2}.
           \label{eq:effectivepotpart}
\end{equation}
From Eq.(\ref{eq:rrest})  for  $q\neq 0$  we  find  the  obvious  rest radius $r_r=0$.   In this  radius the  energy of the  particle is
$E_r=0$. Similarly we  find  the  rest  radii
\begin{equation}
 \frac{r_{r \; (\pm)}}{M}= \frac{ c^2 q^2  - 2 \Sigma G M m^2 (\pm) \sqrt{ c^2q^2\left(c^2q^2
                                  - 8 \Sigma G m^2\right)} }{2 \Sigma m^2 c^2}.
 \label{eq:rrestp}
\end{equation}
 For  the  rest  radius $r_{r\,(\pm)}$  we  have  the  following  values  for  the energy
\begin{equation}
 E_{r \pm}^{(\pm)} = \pm  \frac{m c^2 \sqrt{2 q^2 \left[q^2  -  2 G m^2  (\pm) \sqrt{ q^2 (q^2 - 8 G m)^2}  \right]^3}}{\left[q^2   + \sqrt{q^2 (q^2  - 8 Gm)^2}  \right]^2} ,
  \label{eq:restEnergya}
 \end{equation}
Notice  that for  each  value  of  $r_r$  is possible to have a  particle in rest with  negative energy ($E_{r-}$).  Also notice  that, the rest radii
  are conditioned  by  the  mass  $m$ and  charge $q$ of  the test  particle. For  null  curves  these    are  not defined.  For  time-like  curves
$(\Sigma=c^2)$  the  rest radii for  a charged  particle are restricted  by the  discriminant $q^2 - 8 m^2 G$ in the  formula (\ref{eq:rrestp}). In
fact,  the  rest radii take real positive values  for  all  value  of  $q$ and  $m$ except for  those in the  interval $(-\sqrt{8G}m,  \sqrt{8G}m)$.
Note  that for  the value of  the  charge $q^2 = 8 m^2 G$, we  have  $r_r=3GM/c^2$. For this
value of  charge the  energy of  the charged  test  particle $E_{r +}$  reaches  its  minimum  value which  is $+(3\sqrt{6}/8) m c^2$ whereas
  the  energy  $E_{r -}$  reaches its maximum  value,  which  is $-(3\sqrt{6}/8) m c^2$. Additionally, the  angular momentum and  the  energy for  a  circular  
  orbit with radius $r_c$ are  given  by
\begin{equation}
 L_{c\;\pm} = \frac{q \sqrt{G}M}{c} \mp  \frac{(c^2 r_c + GM)m}{c^2}
                     \sqrt{ \frac{\Sigma G M}{c^2 r_c - GM}}
                       \label{eq:Lco_p}
\end{equation}
and
\begin{equation}
     E_{c\;\pm} =\pm \frac{ m c^4  }
                         { \left(c^2  r_c + GM \right)} \sqrt{\frac{\Sigma r_c^{3}}{{c^2} r_c - {G M}} }
                         \label{eq:Eco_p}
\end{equation}
respectively. From  Eq.(\ref{eq:Lco_p}) and (\ref{eq:Eco_p}) we  conclude   that  in  order  to  have   a  time-like  circular  orbit the  charged
particle must  be placed in  a  radius $r_c > GM/c^2$.  It is  easy  to note that for  the  radius  $r_c= 3GM/c^2$ the  angular  momentum  of  the
particle  is  $L_c^{\pm} = q\sqrt{G}M/c  \mp \sqrt{2}/2GMm/c$. Notice that  for this  radius and the particular  values of  the  charge
$q=2\sqrt{2G}m$  and $q=-2\sqrt{2G}m$ we  have  the  corresponding values  for  the  angular  momentum  $L_c^{\pm} = (1 \mp 1)\sqrt{2}GMm/(2c)$ and
$L_c^{\pm} = (- 1 \mp 1)\sqrt{2}GMm/(2c)$, respectively.  Accordingly,  for this  radius  and  the   charge $q=2\sqrt{2G}m$ we  have  $L^+_c=L_r^+ =0$. 
Similarly, for  this  radius  and  the  charge $q=-2\sqrt{2G}m$  we  have $L^-_c=L_r^- =0$ . In these  cases  the  value  of  the  energy  of  the particle
 is $E_{c\pm} = E_{r\pm}= \pm(3\sqrt{6}/8)mc^2$. Moreover, from the  solution  of   the  equation $d^2\Phi/dr^2=0$ for  $L$
 and Eq.(\ref{eq:Lco_p})
we  find  that   the
last  stable circular  orbit satisfies the  very  simple   equation
$r (c^2 r - 3 GM) =0.$
Therefore the  last  stable circular  orbit  occurs in  $r_{lsco}=3GM/c^2$ and  the  angular momentum  of  the  last stable  circular  orbit  is given
by
 $c^2 L_{lsco\;\pm} = qc \sqrt{G}M \pm  2\sqrt{2 \Sigma } G M m $
As  we  can see, the  angular momentum and   the  energy    are conditioned  by  the  mass  $m$ and  charge $q$ of  the test  particle. As we  note,
for space-like  curves   the  angular momentum of the last  stable circular  orbit it is  not  defined,  whereas  for  a  null  curves the  angular
momentum of the last  stable circular  orbit is  $L_{lsco\;\pm} = {q \sqrt{G}M}/{c}$.  The  angular  momentum  of  a  charged  particle in a
time-like  circular  stable  orbit  is  given  by $ L_{lsco\;\pm} = (q \sqrt{G}  \pm   2\sqrt{2  } G m)M /c$.  Accordingly  we  can  see  that if  the
charge  of  the  particle is $q=2\sqrt{2G}m$  then $L_{lsco-}=L_{c+}=L_{r}=0$. Analogously, if $q=-2\sqrt{2G}m$  then $L_{lsco+}=L_{c-}=L_{r}=0$.
Then we  deduce  that the  particle  can  be  placed  in  a last  stable  circular  orbit of  radius $r = 3 GM/c^2$  for  some arbitrary  values  of  the charge. 
Whereas  the  particle  will  be in  rest in  a  stable  position  $r = 3 GM/c^2$   if  the value of  the  charge is $q=\pm2\sqrt{2G}m$.

The energy   of  the charged particle can  be  written  in terms  of  the its  effective potential $\Phi$  as (see Eq. (\ref{eq:circularconditios}))
$E=\pm m c \sqrt{\Phi}$,  with  $\Phi$ given  by Eq.(\ref{eq:effectivepotpart}).  At  the  infinitum,  the
effective potential (energy)  tends to a  constant.  As  we know,  the  radius  for  the  last stable   orbit  is  given  by $c^2 r = 3 GM$
It  is  possible  then,  for  this  radius and value of the  charge   $q/m=2\sqrt{2G}$,  to  have  a particle a  particle  in rest  or  in an orbit  
with angular  momentum $4\sqrt{2}G/c$ (``clockwise'' motion). A  similar  result is obtained  if the  charge  is  negative. The  angular  momentum  
 in such  case corresponds to  a charged  particle  with angular momentum $-4\sqrt{2}G/c$ (``counterclockwise'' motion). 
 In the  classical  radius $r=3GM/c^2$  circular  orbits  exist with  zero  angular  momentum.  
That phenomena is  interpreted as   a consequence of  the repulsive  force generated by the  charge  distribution. As the  space-time  here  considered 
 is  free  of  singularities we conclude that  this  phenomena  is not  exclusive to the  case  of  naked  singularities.  

\section{The perihelion advance in a  conformastatic magnetized  spacetime}
To  study the  problem of  the  perihelion advance  of a test  charged  particle  in a  conformastatic spacetime  in  presence of  a magnetic  field
we  start  of  the equation  (\ref{eq:circuprecess}). We  restrict  the  motion of  the  particle  to  the plane $z=0$. Then  we  have  the  expression
\begin{equation}
   \frac{d^2u}{d\varphi^2}+  u^2 = F(u), \quad u=1/r,
   \label{eq:orbitu}
\end{equation}
where
\begin{equation}\label{eq:explicF}
  F(u) \equiv \frac{1}{2} \frac{d G}{du},\qquad
 G(u) \equiv \frac {1}   {\left(1 -   \frac{q A_{\varphi}} {c L}  \right)^2}
                \left[  \frac{E^2}{c^2L^2} (1 - U )^4 - \frac{\Sigma m^2}{L^2}(1 -U)^2 \right].
\end{equation}
 Accordingly,  by  following  a  procedure in the same  way  as  in  Ref.~\refcite{Harko2010},  we  have  for the  resulting perihelion advance
   $\delta \varphi = \pi ({dF}/{du})_{u=u_0},$
  where  $u_0$ is  the radius of  a nearly  circular  orbit,  which is  given  by  the  roots  of the  equation $F(u_0)=u_0$.
By  inserting  Eq.(\ref{eq:Upotential}) into  Eq.(\ref{eq:explicF}) we obtain  for $F(u)$
\begin{equation}
F(u)= {\left[  \frac{2E^2 GM}{c^4 L^2}    \left( 1 + \frac{G M}{c^2} u \right)^3
                        - \frac{\Sigma m^2 G M }{c^2 L^2}  \left( 1 + \frac{G Mu}{c^2}  \right)
                         \right]}{\left( 1  - \frac{q \sqrt{G} M}{c L}  \right)^{-2}},
\nonumber
\end{equation}
accordingly,   we find that the perihelion advance in the orbit of the
test particles  in this  kind  of  spacetime is $\delta{\varphi} = \pi ( \psi_0     - k_2^2 )/ Q^2$, where
\begin{equation*}
 \psi_0 \equiv \frac{\left[ 6\left(Q^2 + k_2^2\right)
                  + \left[ 54Q^2k_1 \left( -1 + \sqrt{1 -\frac{6\left(Q^2 +
k_2^2\right)^3}{81Q^4k_1^2}  }    \right) \right]^{2/3}      \right]^2}
{ 6\left[ 54Q^2k_1 \left( -1 + \sqrt{1 -\frac{6\left(Q^2 +
k_2^2\right)^3}{81Q^4k_1^2}  }    \right) \right]^{2/3}  },
\end{equation*}
with
\begin{eqnarray*}
 K_1^2=\frac{E^2G^2M^2}{c^6L^2},\quad
 K_2^2=\frac{\Sigma m^2 G^2M^2}{c^4L^2},\quad \text{and} \quad
 Q^2= \left(1 - \frac{q\sqrt{G}M}{cL}  \right)^2.
\end{eqnarray*}
Notice  that when $q=0$ we  get  the  case in which  $\delta\varphi$ describe  the  perihelion  advance of  a
neutral  particle. As we  recover  the  extreme Reissner-Nordstr\"{o}m solution in  its  standard   form  by  writing  the metric in spherical  polar
coordinates  and  replacing   $R$ by  $R - GM/c^2$ in  the  metric then is  posible  to perform  a  comparison between  the  results presented here
and  the  corresponding results  presented  in  another works.  In  addition, along  this  work  we  have  used the CGS units,  in  such  a way that  
 a comparison  with the  observational  data  can be  applied.


\end{document}